\documentstyle[aps,prl,epsf,floats]{revtex}

\begin{document}
\draft

\title{Scaling properties of the cluster distribution of a critical nonequilibrium model}

\author{Marta Chaves and Maria Augusta Santos}
\address{
Departamento de F\'\i sica and Centro de F\'\i sica do Porto, 
Faculdade de Ci\^encias, Universidade do Porto\\
Rua do Campo Alegre 687 -- 4150 Porto -- Portugal}

\date{\today}
\maketitle

\begin{abstract}
A geometric approach to critical fluctuations of a nonequilibrium 
model is reported. The two-dimensional majority vote model was 
investigated by Monte Carlo simulations on square lattices of 
various sizes and a detailed scaling analysis of cluster statistical 
and geometric properties was performed. The cluster distribution 
exponents and fractal dimension
 were found to be the same as those of the (two-dimensional) Ising
 model. This result, which cannot be derived purely from the known bulk 
critical behaviour, widens our knowledge about the range of validity
 of the Ising universality class for nonequilibrium systems.   

\end{abstract}


\section{Introduction}

Nonequilibrium models, defined by a stochastic evolution rule, 
often develop long range correlations in space and time.
 As a consequence of the absence of the microscopic reversibility 
constraint, critical phenomena in those systems exhibit a wider 
range of universality classes than in the usual thermal
phase transitions. A lot of research activity in nonequilibrium 
systems in recent years was concerned with the characterization of steady states critical behaviour and also with some 
time-dependent properties \cite{marrodick}.

The relation between the singularities in macroscopic
 quantities and the corresponding internal cluster structure in 
critical
 equilibrium systems received considerable attention in 
the seventies and 
eighties [2-8]. 
A scaling theory for the cluster distribution was presented
 \cite{binder}, involving the standard critical indices plus
 a new exponent describing the compactness of the clusters. Geometric
clusters were however found to be too large to describe correlations
 between spins. The concept of critical clusters (also called physical
 clusters or droplets) was introduced
\cite{coniglio80} and is the basis of fast algorithms of thermal
equilibration \cite{sw}.

A number of nonequilibrium models have critical behaviour 
characterized by equilibrium Ising exponents [10-20]; 
it is however not known if their structure on the 'microscopic' 
level is also Ising-like - as Alonso et al \cite{alonso} have found for a nonequilibrium random field model - or rather a distinct one. The 
influence
 of the dynamics on the system steady state geometric properties 
is yet to be clarified.

 Cluster-flipping rules have also
 been used to propose new nonequilibrium models and universality
 classes \cite{grass}. Nonequilibrium models are {\em defined} by
 a particular, usually local, dynamic rule - modifying this rule 
often leads asymptotically to a 
distinct steady state. Since local dynamics suffer from critical 
slowing down, it is desirable to devise alternative (faster!)
 dynamics preserving the statistical properties of a particular 
steady state. This
 raises the interesting question of how to relate geometric to physical clusters 
for a particular nonequilibrium model. As a first step to answer these questions,
 we have analysed the cluster structure of a nonequilibrium 
model that belongs to the two-dimensional kinetic Ising universality 
class, the $2d$ majority vote model \cite{liggett,mario,mendes}.
The results are compared with those obtained in the Ising case.

\section{Model and Formalism}

'Spin' variables ({\em voters}) $\{ \sigma_i \} = \pm 1$ are defined 
on the sites of a square lattice and evolve in time by single-spin flips
with a probability $W_i$ given by

\begin{equation}
W_i = \frac{1}{2} \left[ 1 - \sigma_i (1-2q) S \left( \sum_{\delta} \sigma_{i+\delta}\right) \right]
\end{equation}

where $S(x) = sign(x)$ if $x\neq 0$, $S(0)=0$  and the sum is over
 nearest neighbours of $\sigma_i$; $q$ is the control parameter and 
we will retrict to $0\leq q \leq 0.5$ (alignment with neighbours is 
favoured). At $q = q_c = 0.075$ (for a square lattice), the system
undergoes a nonequilibrium phase transition from a ferromagnetic 
(low $q$) to a paramagnetic (higher $q$) state; the (static
 and dynamic) critical behaviour is Ising-like \cite{GJY,mario,mendes}.

In what follows, a cluster is a connected set of spins such that
 each spin has at least one nearest neighbour in the same state.
 The number of spins in the cluster is the cluster size (or mass)
 $s$; spins with at least one neighbour with the opposite orientation 
belong to the cluster perimeter. By analogy with the Ising case, one
 expects that at criticality there are non-spanning clusters of all 
sizes, i.e. the cluster size distribution $n_s$ (average number of
 clusters of size $s$ per lattice site) has power-law form
 \cite{binder}

\begin{equation}
n_s \sim  s^{-\tau}
\end{equation}

The average perimeter of $s$-size clusters (per spin) $P_s$
 scales as

\begin{equation}
P_s \sim s^{\sigma}
\end{equation}

where $\sigma$ measures the degree of compactness of the clusters
 and varies between $1$ (ramified clusters) and $1-\frac{1}{d}$ 
(compact clusters in $d$ dimensions). 
Corrections to scaling are expected for $q$ slightly away from $q_c$
 and take the form
\begin{equation}
n_s(s,\epsilon) \sim  s^{-\tau}f(s^y\epsilon)
\end{equation}

\begin{equation}
P_s(s,\epsilon) \sim  s^{\sigma}g(s^y\epsilon)
\end{equation}

with $\epsilon = \frac{q_c-q}{q_c}$. Formulas above are valid 
for $1 \ll s \ll L^d$, where $L$ is the system linear size. Assuming
 expressions (2)-(5) hold for the present model, it is straightforward
 to see that, as in the equilibrium case \cite{binder,cambier},
 the distribution exponents $\tau$, $\sigma$
and $y$ are related to the usual critical indices by
\begin{equation} 
\beta = \frac{\tau-2}{y}
\end{equation}

\begin{equation}
\alpha = 1-\frac{\tau-\sigma-1}{y}
\end{equation}

so one additional exponent is sufficient to describe the scaling 
properties of the non-spanning clusters. For an Ising system the
spanning cluster is compact in the ferromagnetic phase but becomes
 a fractal (fractal dimension $D$) at the critical point. In 
contrast to percolation, where $D$ is related to the euclidean
 dimension $d$ through critical exponents of the same model
 \cite{stauffer},  
for Ising or Potts models it was shown \cite{stella,vand89,vand92,marko}
 that $D$ is independent from the corresponding bulk exponents.
It can be obtained from the second moment of the (finite)
 cluster distribution
(the percolative susceptibility)\cite{vand89} 

\begin{equation}
\chi_p = \sum_{s}s^2 n_s \sim \epsilon ^{d-2D}
\end{equation}

or, for a finite system at criticality,

\begin{equation}
\chi_p(q_c, L) \sim L^{-d+2D}.
\end{equation}

A direct measurement of the fractal dimension $D$ of the 'infinite' 
cluster is made  
by counting the mass within a distance $r$ from its
 center of mass
  
\begin{equation}
s(r) \sim r^D.
\end{equation}

or, alternatively, by using the relation between  average 
size of the 'infinite' cluster $s_{\infty}$ and lattice side $L$ 
 
\begin{equation}
s_{\infty}\sim L^D.
\end{equation}

Assuming that the finite clusters have the same fractal dimension \cite{cambier,stauffer,marko,meo}, 
the dependence of cluster size on mean 
radius of gyration $R_s$ also yields $D$

\begin{equation}
s \sim R_s^{D}.
\end{equation}

\section{Monte Carlo Simulations and Results}

We have performed simulations of the model described in the previous
 section on square lattices of $32^2$, $40^2$, $64^2$, $128^2$
 and $200^2$ sites
 with periodic boundary conditions for several values of
 $q \leq q_c$. After reaching a stationary state for each value 
of $q$, the system evolution was monitored for up to 
$1.2*10^6$ MCS; configurations were analysed every 200
 MCS. Clusters of either 'aligned' or 'reversed' spins 
(compared to the dominant sample orientation) were identified
 and counted separately, together with the respective perimeters.
 Double logarithmic 
plots of $n_s(s)$ and $P_s(s)$ - for reversed spins - are displayed in
 figures 1 and 2, 
respectively, for several $q$ values; the slopes of the 
corresponding linear regions are consistent with the Ising 
exponents $\tau = 2.05$ and $\sigma = 0.68$. From fig.2 we notice 
that the perimeter scaling law (3) is only valid for intermediate size clusters, larger
 clusters being more ramified than small ones, as found in the Ising 
case \cite{cambier}. A better estimate
 of $\tau$, together with the correction to scaling exponent 
$y$, are obtained from the best scaling fit to the data of
 the function \cite{cambier} 

\begin{equation}
F(\epsilon S ^y) \equiv \epsilon^{1/y} \sum_{s=1}^{S} s^\tau n_s(s,\epsilon)
\end{equation}

as shown in fig.3. This data collapse yields $\tau = 2.05\pm 0.005$ and $y=0.445 \pm0.01 $.

The study of the fractal dimension of finite (inverted) clusters 
at
 $q_c$ is presented in fig.4 for the largest system size ($L=200$). 
In 
order to reduce statistical scatter, the results were collected
 in bins of 
growing size (bin width increases by a factor of 1.5 for each 
successive interval). The 
best fit to the data yields $D=1.938 \pm 0.003$; smaller systems
 produced slightly higher values, so we estimate $D=1.945\pm0.01$. This is in agreement with
the value 
$D=1.95 \pm 0.02$ obtained from the finite size relation (9) 
for $L=32,40,64,128$ and coincides, within error, with the exact 
Ising value $D=1.947$ \cite{stella}.

Fig.5 shows a direct evaluation of the fractal
 dimension of the percolating 
cluster at $q_c$. The best fit for the $L = 200$ system gives 
$D = 1.970 \pm  0.003$ to be compared with $D = 1.968$ obtained from
 relation
 (11) for $L = 64, 128$ and $200$. This slightly higher value
 of $D$ 
obtained from the infinite clusters data is interpreted as a 
finite 
size effect: even for the largest lattice, the magnetization
 is
 finite at $q_c$ and the percolating cluster was allways
found to
 have the dominant orientation of the sample. Hence, we were
 probably not 
probing the {\em incipient infinite cluster} but rather a more
 compact
 object.

\section{Conclusion}
We have performed a detailed numerical study of the statistical 
properties of geometric clusters close to the critical point of a two-dimensional majority vote model.
Scaling was confirmed with distribution exponents
$\tau=2.05\pm 0.005$,  $\sigma=0.68\pm0.01$ and  $y=0.445\pm
0.01$, indistinguishable
 from their equilibrium Ising values,
 as reported in reference \cite{alonso} for another model in 
the same universality class.
The fractal dimension associated with the cluster distribution at
 criticality $D=1.945\pm0.01$ was also found to coincide with its Ising value
 $D = \frac {187}{96}$. A direct determination of $D$ from the
 geometric properties of the 'infinite' cluster
 produced a slightly higher value, which we believe is due to
 the finite system size. The percolative susceptibility
 diverges 
with an exponent $\gamma_p \equiv -2+2D=1.90$ larger than the
 Ising
value $\gamma = 1.75$, indicating that, as expected, purely 
geometric clusters
 do not coincide with critical ones. Coniglio and Klein's rule \cite{coniglio80} to reduce geometric to physical clusters - 
introducing 'active bonds' with a temperature-dependent probability - 
is not easily adaptable to this system. The difficulty to relate the probability for an 'active bond'  
with the noise parameter $q$ is due to the non-hamiltonian nature
of the model. The study of the 
majority vote phase transition as a percolative transition of
 (suitably defined) physical clusters is a challenge for 
future work.

\acknowledgements
We thank J M B Oliveira for his help in the implementation of the
 cluster counting algorithm and J F Mendes for usefull discussions.
This work was partially financed by Praxis XXI (Portugal) within
 project PRAXIS/2/2.1/Fis/299/94; M C has beneficted from a junior
 research grant (BIC) from Praxis XXI.

%=========================================================================

\newpage
\section{Figure Captions}

Fig.1 - Double logarithmic 
plot of the cluster size distribution. From bottom to top $q=0.070, 0.071, 0.072, 0.074$ and $0.0747$
(successive curves have been shifted one unit up for clarity). 
The straight 
lines have slope $-2.05$. Lattice size $L=200$; averages over 
5000 or 6000 ($q=0.0747$) samples.\\

Fig.2 - Log-log plot of the perimeter length $P_s$ as a function of
 size $s$.
Same conditions as in fig.1. The dotted lines have slope $0.68$. The 
inset shows  a zoom of the linear region for the five $q$ values; the
straight line is a least-square fit and has slope $0.683$. \\

Fig.3 - Scaling function $F$ versus $\epsilon S^y$ for $\tau=2.05$ and $y=0.445$ (see text). The arrows 
indicate end of data for (left to right) $q=0.0747,
0.074, 0.072, 0.071$ and $0.070$. Lattice size $L=200$.  \\

Fig.4 - Cluster size v.s. mean radius of gyration in double 
logarithmic
scale for $L=200$ and $q=q_c$; data binned as described in
 the text. The
 best fit line has slope 1.938. Averages over 5000 samples.\\

Fig.5 - Mass within distance $r$ from center of mass, $s_r \equiv s(r)$, 
for percolating clusters at $q_c$. Averages over 5000 samples.
System sizes $L=64(\Box), 128(\bigcirc)$
 and $200(+)$. The straight line has slope $D=1.970$. 

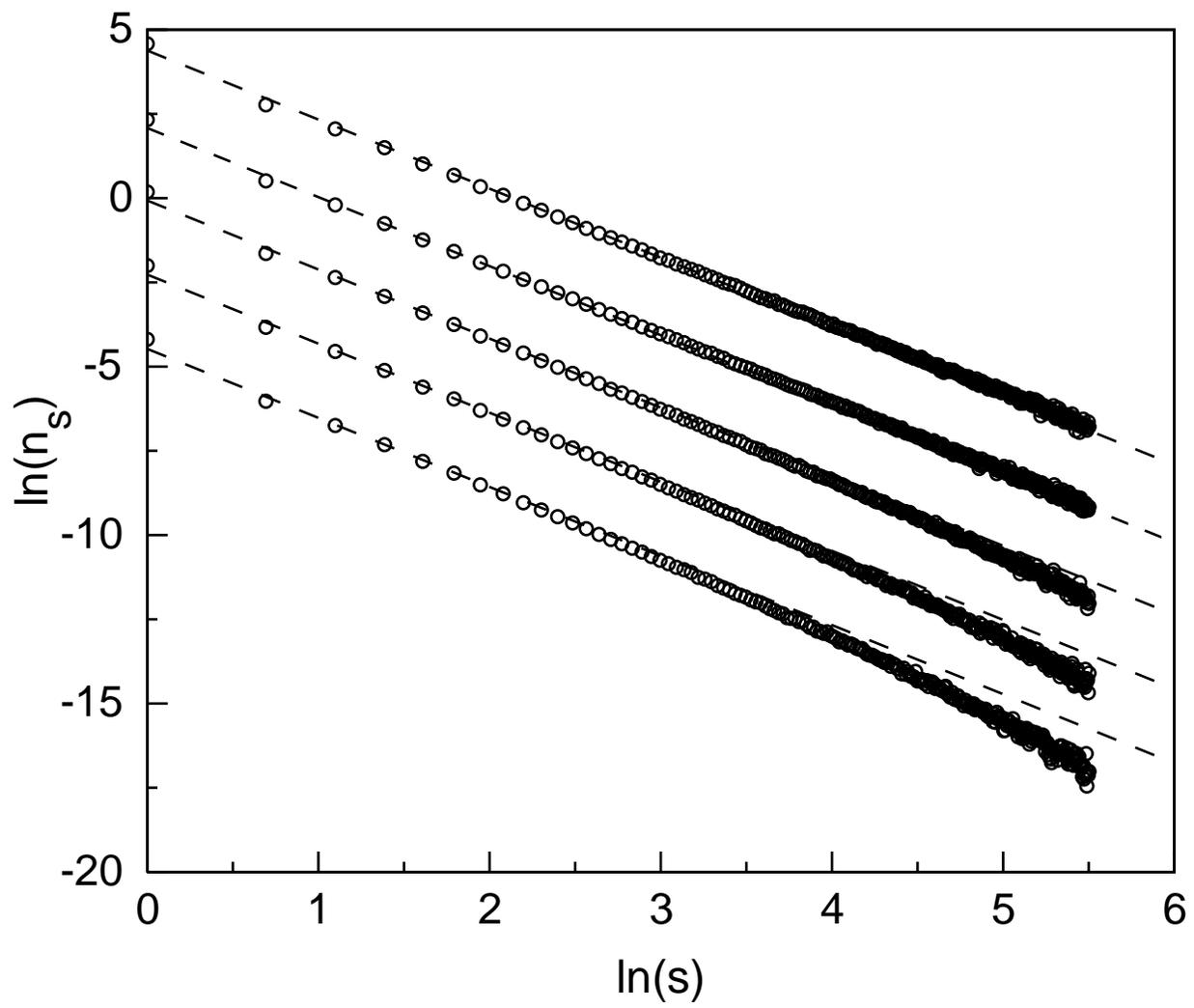
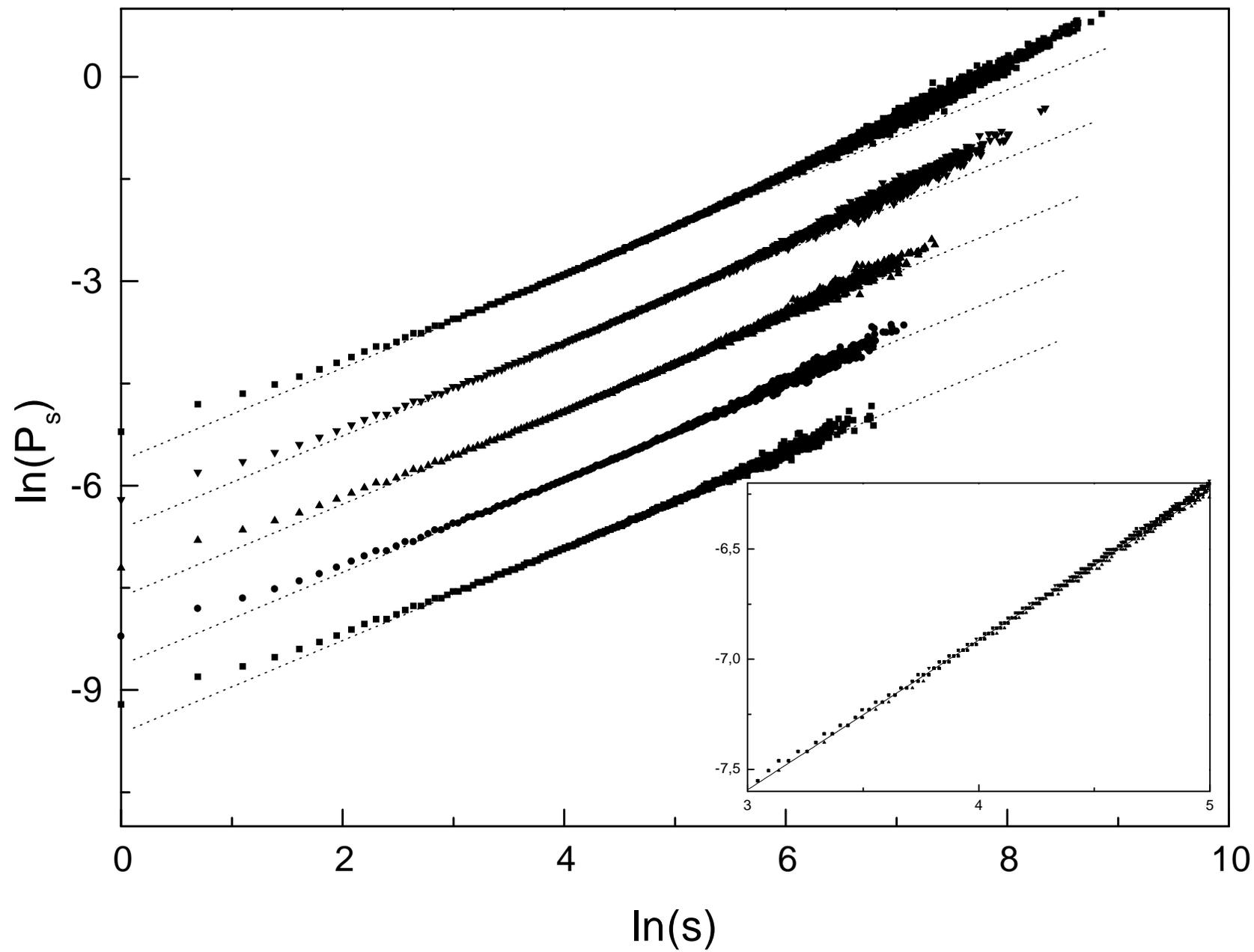
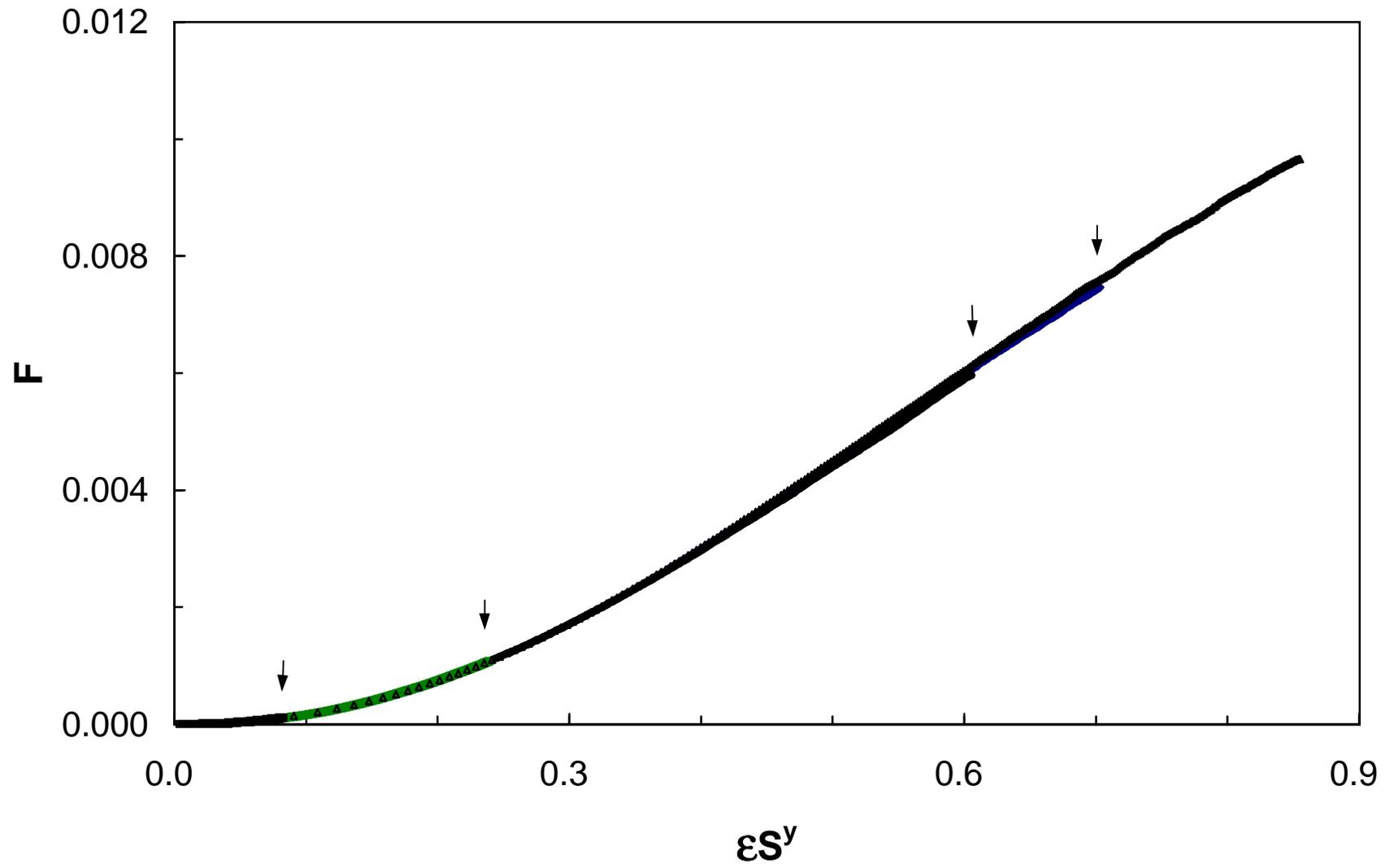
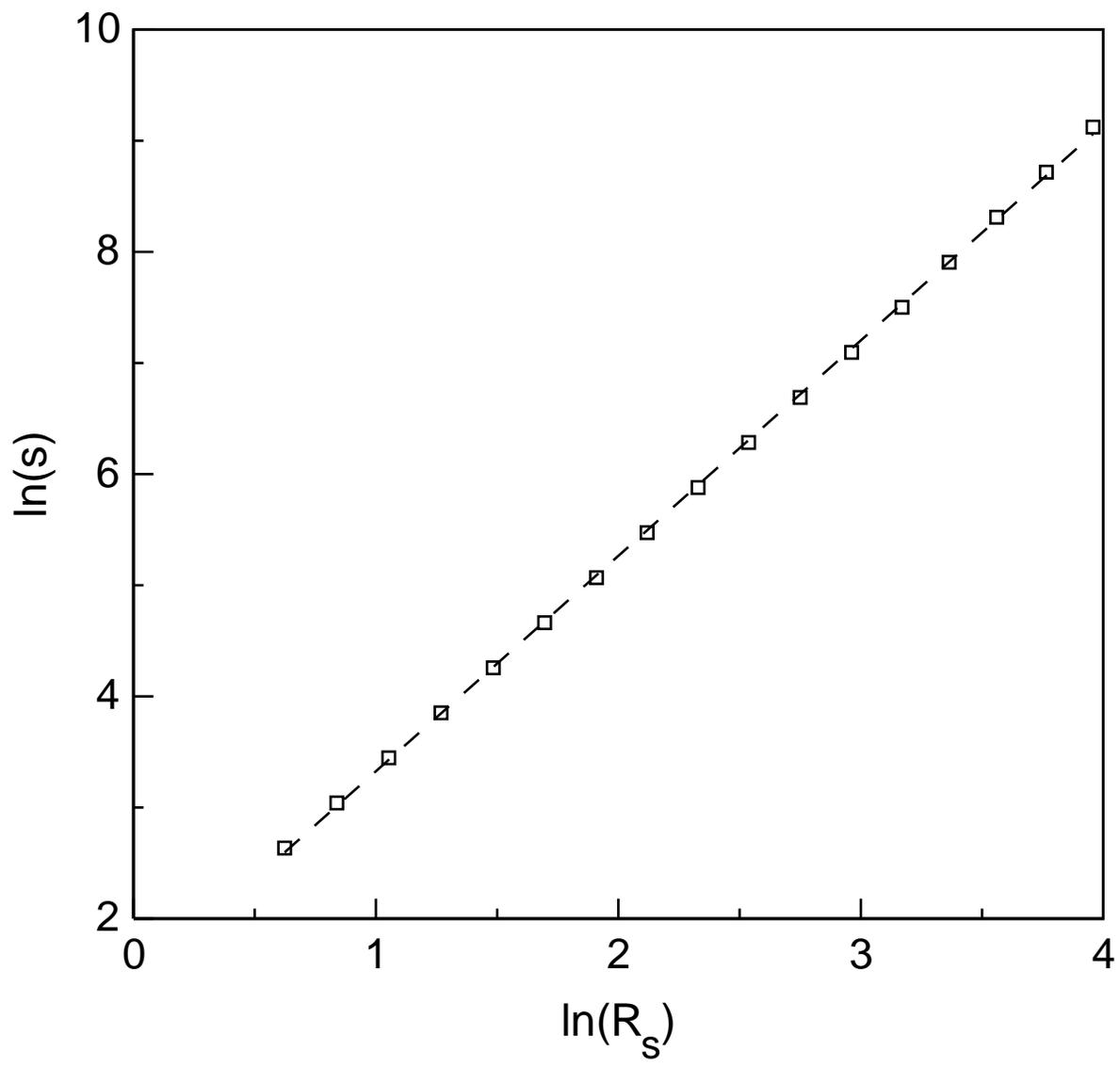
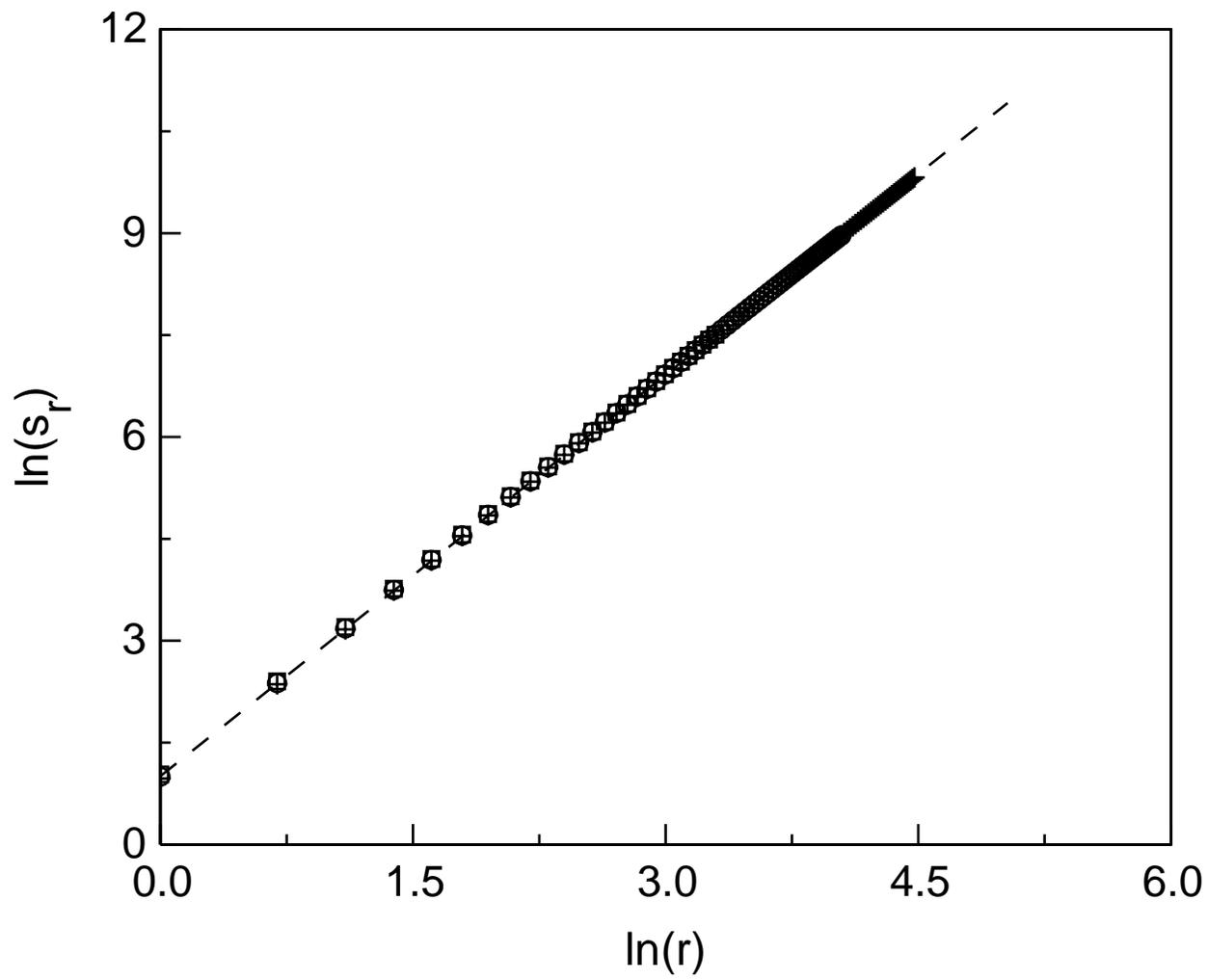
\end{document}